\documentstyle[12pt,epsfig]{article}
\begin{document}
\begin{quote}
\raggedleft
hep-ph/yymmxxx \\
PM/96-16 \\
June 1996
\end{quote}
\begin{center}
{\bf \Large Radiative corrections to $e^+e^-\to H^+ H^-$~\footnote{
To appear in {\it Proceedings of the Physics with $e^+e^-$ Linear 
Colliders Workshop}, 
Annecy -- Gran Sasso -- Hamburg 1995, ed. P. Zerwas}}

\vspace*{0.3cm}

A. Arhrib$^{a,b}$ and G. Moultaka$^a$ \\
\vspace*{5mm}
$^a$ Physique Math\'ematique et Th\'eorique, E.S.A. du 
CNRS N$^o$ 5032, \\ 
Universit\'e Montpellier II, F-34095 Montpellier France\\
\vspace{2mm}
$^b$ L.P.T.N., Facult\'e des Sciences Semlalia,
B.P. S15, Marrakesh, Morocco
\end{center}
\baselineskip=18pt


\begin{abstract}
We study the 1-loop corrections to the charged Higgs production both in the 
Minimal Supersymmetric Standard Model (MSSM) and in a more general type II 
two-Higgs-doublet model (THDM-II). 
We consider the full set of corrections (including 
soft photon contributions as well as box diagrams), and define a 
parametrization that allows a comparison between the two models.
Besides the soft photon radiation there can be prominent model-dependent 
effects.    
\end{abstract}
\newpage
\begin{center}
{\bf \Large Radiative corrections to $e^+e^-\to H^+ H^-$}

\vspace*{0.3cm}

A. Arhrib$^{a,b}$ and G. Moultaka$^a$ \\
\vspace*{5mm}
$^a$ Physique Math\'ematique et Th\'eorique, E.S.A. du 
CNRS N$^o$ 5032, \\ 
Universit\'e Montpellier II, F-34095 Montpellier France\\
\vspace{2mm}
$^b$ L.P.T.N., Facult\'e des Sciences Semlalia,
B.P. S15, Marrakesh, Morocco
\end{center}
\baselineskip=18pt


\begin{abstract}
We study the 1-loop corrections to the charged Higgs production both in the 
Minimal Supersymmetric Standard Model (MSSM) and in a more general type II 
two-Higgs-doublet model (THDM-II). 
We consider the full set of corrections (including 
soft photon contributions as well as box diagrams), and define a 
parametrization that allows a comparison between the two models.
Besides the soft photon radiation there can be prominent model-dependent 
effects.    
\end{abstract}

\section{Introduction}

In contrast to hadronic machines, a high energy $e^+ e^-$ collider in the TeV 
range will be a rather unique place to discover and study charged higgses 
in a clean environment. 
These would be produced either in pairs \cite{komamiya}, our main concern here,
or in (rare) production in association with $W^\pm$. 
It was first found in \cite{ACM}  that  loop corrections from matter fermions 
and their susy partners (mainly the $(t,b), (\tilde{b}, \tilde{t}) $ sector), 
are likely to change the tree-level result at $\sqrt{s} = 500 $ GeV 
\cite{komamiya,tree-H}, by as much as $10\%$  dip in the cross-section.
The effect could even lie between  $-25\%$ and $25\%$ and perhaps grow out of 
perturbative control, though in a realistic range of the model-parameters. 
Such a sensitivity to loop effects appears to be related to 
the fact that at tree-level the $\gamma$ and $Z$ 
mediated process is exclusively controlled by $U_B(1) \times U_{W_3}(1)$ gauge 
invariance and thus knows nothing about the structure of the non-standard 
extension whatsoever.

It is thus important in a first step to go beyond \cite{ACM} and study the full
1-loop structure, in order to understand the sensitivity to the various 
model-parameters and the origin of large effects, whether in the MSSM or in
a general THDM-II. 
In a second step one should perform a renormalization group analyses, 
eventually taking into account constraints from SUGRA-GUT models, and 
ask whether the allowed regions of the model-parameters could still lead 
to large or even uncontrollable loop effects in charged Higgs pair production.   
    
In the present study  we resume the first of these steps and improve 
upon the results of ref \cite{ACM} by including:

{\bf a)} the complete Higgs sector contributions (self-energies, vertices 
and boxes);

{\bf b)} the  infrared part, including initial and final soft photon 
radiation as well as $\gamma \gamma$ and $\gamma Z$ boxes;

{\bf c)} The complete set of charginos/neutralinos/$\tilde{e}$/$\tilde{\nu}$ 
 box diagrams;

We performed the analysis in the on-shell renormalization scheme defined in  
\cite{BHS}. Since at tree-level only standard model couplings appear
in $e^+e^-\to H^+ H^-$, we need only to supplement the above scheme with the
wave function and mass renormalization conditions of the charged Higgs field.
It is also convenient to take the charged Higgs mass $M_{H^\pm}$, 
rather than that of the CP-odd scalar $M_{A^0}$, as the input on-shell physical 
mass (see \cite{ACM} for more details ).
It will turn out that besides the sensitivity to the heavy quark-squark sector,
there are on one hand model-independent large effects from the soft photon radiation,
and on the other possibly important effects in the vertex corrections involving 
the purely Higgs sector. The latter case occurs when deviations from the 
tree-level MSSM triple-self-couplings are allowed. One of the issues will be the 
possibility of signing a clear difference between the MSSM and a non-supersymmetric 
THDM-II.

\section{Effective Parametrization}

In this section we outline a general parametrization describing deviations 
from the tree-level Higgs self-couplings as given in the MSSM. 
Let us recall first that the most general gauge invariant scalar 
potential (corresponding to CP-invariant THDM-II) is given by
(see for instance ref. \cite{Gunion}):
\begin{eqnarray}
& & V(\Phi_{1},\Phi_{2})=\lambda_{1} (\Phi_{1}^+\Phi_{1}-v_{1}^2)^2
+\lambda_{2} (\Phi_{2}^+\Phi_{2}-v_{2}^2)^2+
                    \lambda_{3}((\Phi_{1}^+\Phi_{1}-v_{1}^2)+(\Phi_{2}^+
\Phi_{2}-v_{2}^2))^2                 \nonumber\\ [0.2cm]
                    & &+\lambda_{4}((\Phi_{1}^+\Phi_{1})(\Phi_{2}^+
\Phi_{2})-(\Phi_{1}^+\Phi_{2})(\Phi_{2}^+\Phi_{1}))+
                    \lambda_{5} (Re(\Phi^+_{1}\Phi_{2})
-v_{1}v_{2})^2+\lambda_{6}Im(\Phi^+_{1}\Phi_{2})^2
\label{pot}
\end{eqnarray}
where all $\lambda_i$ are real and $v_1$, $v_2$ denote the vacuum 
expectation values of the two Higgs fields. 
We have also omitted an additive constant in eq. (\ref{pot}).  
Thus one has in general 7 free parameters, i.e. the $\lambda_i$'s and
$\tan \beta(=v_2/v_1)$ , $\sqrt{v_1^2 + v_2^2}$ being fixed by the electroweak
scale. Let us now define 
\begin{eqnarray}
&&\lambda_1= \lambda_2 +\delta_{12} \ \ \ \ \ \ \ \ \ \ ,\ \ \ \ \ \ \ \ \
\lambda_3=  \frac{1}{8}(g^2+g'^2)-\lambda_1 +\delta_{31} \nonumber\\
&& \lambda_4= 2 \lambda_1 -\frac{1}{2}g'^2 + \delta_{41}  \ \ \ \ \ , \
\ \ \ \lambda_5=  -\frac{1}{2}(g^2+g'^2)+ 2 \lambda_1  +
\delta_{51}\\
&&\lambda_6=  -\frac{1}{2}(g^2+g'^2)+ 2 \lambda_1 + \delta_{61}\nonumber
\label{lam}
\end{eqnarray}
where the $\delta_i$'s $\neq0$ in general. It is well known \cite{Gunion} 
that (softly broken) supersymmetry forces the  $\delta_i$'s
to zero. From this follow the usual tree-level relations among the Higgs masses,
$\tan 2\alpha$ and $\tan \beta$, as well as the negative sign of $\sin 2\alpha$.
In this case the triple scalar couplings read (here we only show those entering
the 1-loop corrections in $e^+e^-\to H^+ H^-$): 
\begin{eqnarray}
g_{H^0H^+H^-}^{MSSM}& =& -i g (m_W \cos(\beta - \alpha) - \frac{m_Z}{2 c_w} 
\cos 2 \beta \cos(\beta + \alpha)) \nonumber \\
g_{h^0H^+H^-}^{MSSM}& =& -i g (m_W \sin(\beta - \alpha) + \frac{m_Z}{2 c_w} 
\cos 2 \beta \sin(\beta + \alpha)) \nonumber \\
g_{H^0 H^{\pm}G^{\mp}}&=&
\frac{-i g sin(\beta-\alpha) (m_{H^\pm}^2-m_H^2)}{2 m_W}\nonumber\\
g_{h^0 H^{\pm}G^{\mp}}&=&
\frac{i g cos(\beta-\alpha) (m_{H^\pm}^2-m_h^2)}{2 m_W}\nonumber\\
g_{A^0 H^{\pm}G^{\mp}}&=& 
\mp \frac{m_{H^\pm}^2-m_A^2}{2 m_W} 
\label{gH}
\end{eqnarray} 

As one can easily see, none of the couplings in eq. (\ref{gH}) can be enhanced
at tree-level in the supersymmetric case. In the non supersymmetric case however , 
$\delta_i \neq 0$, there is an important difference between $g_{H^0H^+H^-}$ 
and $g_{h^0H^+H^-}$ on one side, and the three remaining couplings of 
eq. (\ref{gH}) on the other. 
The couplings involving the charged goldstone bosons $G^\pm$
retain their initial form, so that a possible enhancement is due only to deviations
from the supersymmetric Higgs-boson-mass sum rules. In contrast $g_{H^0H^+H^-}$ 
and $g_{h^0H^+H^-}$ will deviate from their forms in eq. (\ref{gH}) through a
non trivial dependence on $\tan \beta$, $m_{H^\pm}$, $\lambda_3$, etc... . 
These aspects will be of importance in what follows.

Starting from eq.(2) it is possible 
to express all couplings and mass relations in terms of the non-vanishing 
$\delta_i$'s, $\tan \beta$ and one of the $\lambda_i$'s. 
(The general expressions are somewhat involved and will be given
elsewhere). Here we would like to insist on a physically interesting 
special case:
It is possible to define a deviation from the MSSM in such a way that {\sl all}
the MSSM tree-level relations\footnote{Strictly speaking one should include the 
known radiative corrections to these relations, however due to our Higgs mass 
renormalization condition they would correspond to higher order effects in 
$e^+e^- \to H^+ H^-$} among the higgs masses, $\tan 2\alpha$ and 
$\tan \beta$ remain valid. 
This occurs provided we require the $\delta_i$'s to satisfy
\begin{eqnarray}
&& \delta_{51}=\frac{2 tan^2 \beta}{1- tan^2\beta} \delta_{12}
\nonumber \\
&& \delta_{61}=\delta_{41}=-2 \sin^2(\beta) \delta_{12} \nonumber \\
&&\delta_{31}= \frac{1}{2} \sin^2(\beta)( 1- \frac{1}{\cos 2\beta})
\delta_{12}  
\label{quasisusy}
\end{eqnarray}
Under such conditions we have only three free parameters which we can choose
as $\tan \beta$, $M_H^\pm$ and $\lambda_3$. The MSSM corresponds then to a 
particular value of $\lambda_3$, $\lambda_3^{MSSM}$ which is a function of 
$\tan \beta$ and $M_H^\pm$ and leads to $\delta_i=0$ as a special case of
eq. (\ref{quasisusy}). For this reason eq. (\ref{quasisusy}) will be dubbed 
quasi-supersymmetric. It can be thought of as corresponding to the situation 
where all the two-Higgs doublet spectrum has been experimentally unraveled 
and found to be consistent with the MSSM Higgs mass sum rules but with no other 
direct evidence for supersymmetry. Although not generic, the quasi-supersymmetric 
parametrization would then be a very useful device in terms of which the tests of the
Higgs self-couplings can be expressed. 
As far as the charged Higgs sector is concerned only $g_{H^0 H^+ H^-}$ and $g_{h^0 H^+ H^-}$ 
can deviate from their supersymmetric values given in eq.(\ref{gH}).  
We give here for illustration their behavior in the large $\tan \beta$ limit (with  
$\lambda_3 \neq  \lambda_3^{MSSM}$ and $m_A > m_Z$
\begin{eqnarray}
g_{H^0H^+H^-}& \sim&  g_{H^0H^+H^-}^{MSSM} - i g m_W (\frac{1}{2 c_w^2}+\frac{m_A^2}{m_W^2}+
\frac{s_w^2}{\pi \alpha} \lambda_3)\tan \beta \nonumber \\
g_{h^0H^+H^-}& \sim&  g_{h^0H^+H^-}^{MSSM} - i g m_W \frac{2 m_A^2}{m_A^2-m_Z^2} 
(\frac{1}{2 c_w^2}+\frac{m_A^2}{m_W^2}+
\frac{s_w^2}{\pi \alpha} \lambda_3) 
\label{largetbeta}
\end{eqnarray}
where $g_{H^0H^+H^-}^{MSSM}$, $g_{h^0H^+H^-}^{MSSM}$, $g_{H^0 H^{\pm}G^{\mp}}$,
$g_{h^0 H^{\pm}G^{\mp}}$ and $g_{A^0 H^{\pm}G^{\mp}}$ are given in 
eq.(\ref{gH}). Note that triple couplings of gauge  bosons to Higgses are 
never enhanced, and quartic couplings enter exclusively one-loop   
diagrams that vanish with $m_e$. It is thus interesting to note that the sensitivity to
large $\tan \beta$ resides exclusively in $g_{H^0H^+H^-}$ of eq. (\ref{largetbeta}),
while $g_{h^0H^+H^-}$ is more sensitive to $m_A$ in this limit. 
A more general investigation of the full Higgs sector can be persued along the same lines.   

\section{Numerical Results}

We now present some numerical results, using the exact expressions for the Higgs self-couplings 
in terms of the 3 free quasi-susy parameters, $\tan \beta$, $m_{H^\pm}$ and $\lambda_3$.
 
In fig.1 we show the percentage contribution to the integrated cross section, for each sector
separately, as a function of $\sqrt{s}$.   
One finds that the Higgs sector contributions can counterbalance those of
the heavy quarks found in \cite{ACM} for large $\tan (\beta)$, but only near threshold
and for $\lambda_3 \neq  \lambda_3^{MSSM}$.  
Far from threshold all of the effects, except for $WW$ boxes, become negative. 
Furthermore the ``neutral'' model-independent contributions, including soft bremsstrahlung,
obtained by adding one photon (or Z) line to the tree diagrams depend loosely on 
$M_{H^\pm}$ or $\sqrt{s}$ and contribute at the level of $-17\%$ for a
soft photon cut $\Delta E_\gamma \sim 0.1 E_{beam}$.
In fig.2 we show (excluding those ``neutral'' contributions) the integrated cross-section 
for two values of $M_{H^\pm}$ and $\tan (\beta)$. In THDM-II the total loop effect increases 
(negatively) with increasing $\tan (\beta)$, the farther one goes from production threshold.  
In the MSSM ($\lambda_3= {\lambda_3}^{MSSM}$ ) the leading effects come exclusively from
the heavy quark-squark sector and the conclusions of \cite{ACM} remain unaltered in this case. 
The $\chi^\pm/ \chi^0/\tilde{e}$/$\tilde{\nu}$ contributions do not exceed a few percent
despite the large number of diagrams. For instance we evaluated the full set of supersymmetric 
boxes, (taking into account fermion-number violation through the rules of ref. \cite{Denner}),
and found that they largely cancel among each other, leading at most to $1-3 \%$
(negative) effect for a wide range of sparticle masses.
Finally we should stress that the large effects in THDM-II are not an artifact
of the quasi-susy parametrization, they are also present even if the tree-level
Higgs mass sum rules are relaxed. This raises the question of whether such 
effects would appear in the MSSM as a result of the running of the 
$\lambda_i$'s \cite{RGimp}. 

We conclude that the charged Higgs sector seems to offer a non-trivial 
structure beyond the tree-level, which can complement the information
gained from the neutral sector. Yet a more elaborate strategy involving
the full-fledged Higgs sector still needs to be worked out.


\begin{figure}[htbp]
\mbox{\epsfig{file=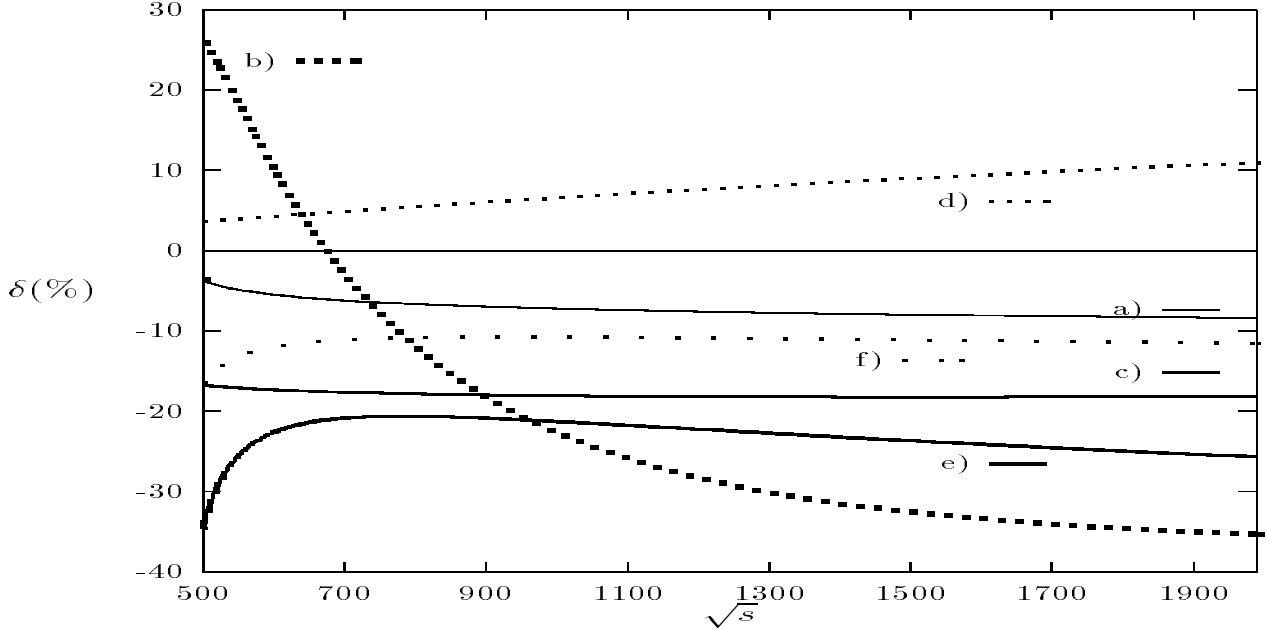,height=8cm,width=12cm,
bbllx=100,bblly=400,
bburx=350,bbury=700}}
\caption{\label{fig1} Contributions in \% to the integrated cross-section in quasi-susy, $\lambda_3= -0.61, M_{H^\pm}=220$; a) Higgs sector, $\tan \beta=2$; b) Higgs sector, $\tan \beta=30$; c) virtual Z, $\gamma$ and soft bremsstrahlung; d) virtual W
boxes; e) matter fermion sector, $m_{top}=180 GeV, \tan \beta=30$; f) same as
e) but with $\tan \beta=2$;}
\end{figure}

\begin{figure}[htbp]
\mbox{\epsfig{file=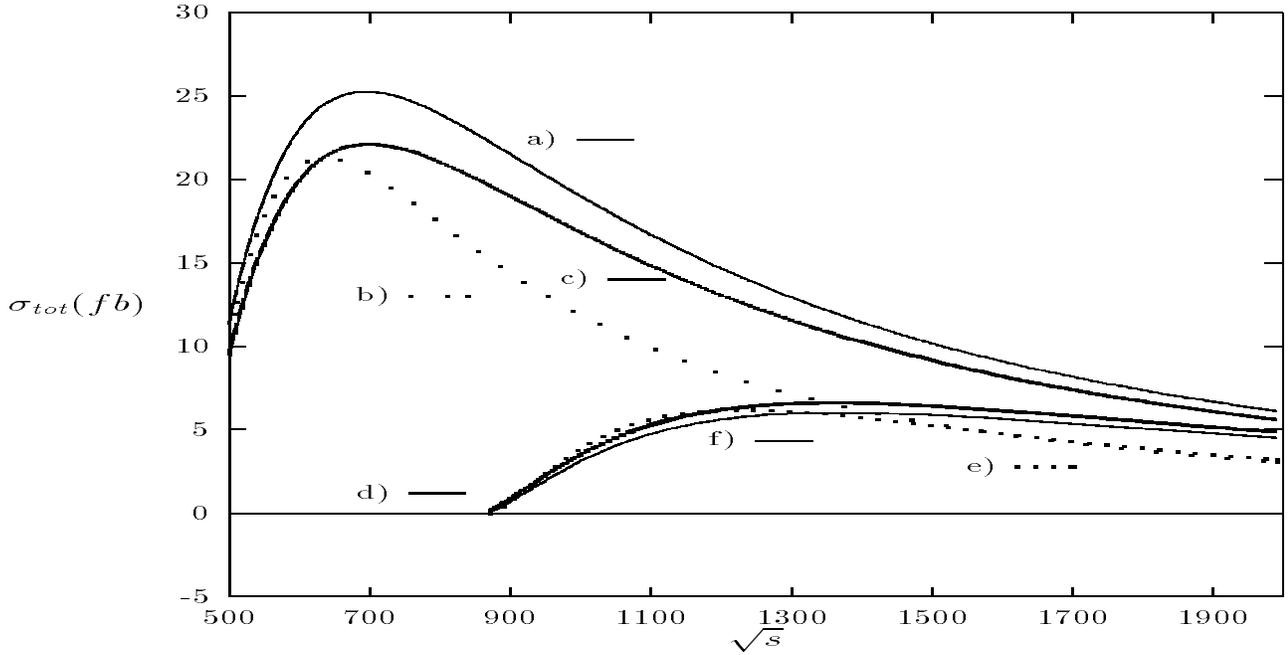,height=8cm,width=12cm,
bbllx=100,bblly=400,
bburx=350,bbury=700}}
\caption{\label{fig2}a) Tree-level, $M_{H^\pm}=220 GeV$; b) quasi-susy, $\lambda_3= -0.61$ (MSSM value $-0.71$), $\tan \beta=30$; c) quasi-susy, $\lambda_3= -0.61, \tan \beta=2$;
d) Tree-level, $M_{H^\pm}=430 GeV$; e) quasi-susy, $\lambda_3= -2.6$ (MSSM value $-2.84$ ), $\tan \beta=30$; f) same as e) but with $\tan \beta=2$; $m_{top}=180 GeV$}
\end{figure}

\end{document}